\documentclass[letterpaper, 10 pt, conference]{ieeeconf}
\pdfoutput=1
\IEEEoverridecommandlockouts                              

\usepackage[utf8]{inputenc}
\usepackage{biblatex}
\usepackage{wrapfig}
\usepackage{times}
\usepackage{lipsum}
\usepackage{amsmath}   
\usepackage{tikz}
\usepackage{graphicx}
\usetikzlibrary{positioning} 
\usepackage{amssymb}	    
\usepackage{amsfonts}	    
\usepackage{mathtools}		
\usepackage{xcolor}
\usepackage{caption}
\usepackage{subcaption}
\usepackage{mathtools}
\usepackage[abs]{overpic} 
\usepackage{soul}
\newcommand{\B}[1]{\if#1\relax\bm{#1}\else\mathbf{#1}\fi}
\newcommand{\R}[1]{\mathrm{#1}}

\newcommand{\BB}[1]{\mathbb{#1}}

\title{External control of a genetic toggle switch via Reinforcement Learning}

\author{Sara Maria Brancato$^1$, Francesco De Lellis$^1$, Davide Salzano$^1$, Giovanni Russo$^{2,*}$, Mario di Bernardo$^{1,*}$
    \thanks{$^{1}$Department of Electrical Engineering and ICT, University of Naples
    Federico II, Italy}
    \thanks{$^{2}$Department of Information and Electrical Engineering and Applied
    Mathematics, University of Salerno, Italy}
    \thanks{$^*$Corresponding authors. mario.dibernardo@unina.it,giovarusso@unisa.it}
}

\date{January 2022}

\begin{document}

\maketitle

\begin{abstract}
We investigate the problem of using a learning-based strategy to stabilize a synthetic toggle switch via an external control approach. To overcome the data efficiency problem that would render the algorithm unfeasible for practical use in synthetic biology, we adopt a sim-to-real paradigm where the policy is learnt via training on a simplified model of the toggle switch and  it is then subsequently exploited to control a more realistic model of the switch parameterized from \emph{in-vivo} experiments. Our \emph{in-silico} experiments confirm the viability of the approach suggesting its potential use for \emph{in-vivo} control implementations.
\end{abstract}

\section{Introduction}
The problem of devising controllers to tame the dynamics of synthetic biological systems is becoming a crucial part of Synthetic Biology \cite{9030209,del2018future} giving rise to the emerging field of {\em cybergenetics} \cite{ruolo2021control}. In the external control paradigm, some phenotype of cells hosted in some environment (typically a turbidostat or a microfluidic chamber) is controlled via an external control action provided to the cells via actuated syringes whose motion is decided by some control strategy implemented in a PC or microcontroller.
Examples of such approaches include those presented in \cite{shannon2020vivo} (see \cite{9030209} for a review of other results from the literature).

Typically, either PI or MPC control strategies have been used to solve this problem (see \cite{guarino2020balancing} for a comparison of different techniques) despite the fact that 
models of the  system under control are typically uncertain and that the sensed data are noisy. Moreover, running experiments is often costly and time-consuming and this can make model identification challenging in practice \cite{del2018future}. 
An alternative approach to address these issues could be to leverage learning-based control methods to learn the policy by directly interacting with the system. A key advantage of these methods is that knowledge of a well calibrated mathematical model is not necessarily required. However, the learning process can be sample inefficient, requiring long times and enough experimental data to learn the policy that could hinder its use in biology \cite{BUSONIU20188,MC1}. 

A promising solution to overcome these problems is the {\em sim-to-real} approach where the control policy is first learnt on simulated environments and  subsequently exported on the real system \cite{james2017transferring}.
However, when this is done, one needs to verify that the {\em sim-to-real} gap can be bridged; i.e., that the policy learnt on the simulated environment can also control the real system.

In this paper, we focus on the problem of stabilizing a synthetic toggle switch about its unstable equilibrium point.
%

Such a problem is a widely adopted benchmark, see e.g. \cite{lugagne2017balancing}, in control applications to synthetic biology, which has been proposed as the equivalent in synthetic biology of of the swinging-up problem of an inverted pendulum in classical control. It is also a problem of practical interest as this bistable system is a fundamental synthetic circuit to endow cells with memory-like features \cite{Hillenbrand2013} or to differentiate mono-strain cultures in different populations \cite{ Laurent1999,salzano2019ratiometric}

Here, we explore the use of  tabular learning to solve the problem adopting an external control setting where the control input is delivered implementing \emph{in-silico} a realistic experimental set-up with actuation and sensing constraints. To overcome the data efficiency problem that would render the algorithm unfeasible for practical use in synthetic biology, we adopt a \textit{sim-to-real} paradigm where the policy is learnt via training on a simplified model of the toggle switch and  it is then  exploited to control a more realistic, higher-dimensional model of the switch parameterized from \emph{in-vivo} experiments in \cite{lugagne2017balancing}. We show via a set of exhaustive \emph{in-silico} experiments that such approach can effectively bridge the \textit{sim-to-real} gap and, therefore, that learning-based strategies can  be used for the control of synthetic biological systems. To  benchmark our results against other state-of-the art algorithms for the toggle switch stabilization, we define and use a set of appropriate control metrics that confirm the effectiveness and viability of the proposed approach.  

\subsubsection*{Related work} the problem of stabilizing the unstable equilibrium of the toggle switch has been recently considered in \cite{lugagne2017balancing} where the regulation problem is solved via model-based open loop control, which exhibits poor robustness with respect to parameter variations. We also recall here \cite{fiore2018analysis}, where a mathematical model describing the dynamics of the toggle switch when this is subject to periodic stimuli is presented. Based on this model, in \cite{guarino2020balancing} a closed-loop proportional integral pulse width modulation controller (PI-PWM) and a MPC strategy are proposed. Although exhibiting good performance and robustness properties, the PI-PWM requires the offline calibration of the control gains via an exhaustive search, while the MPC requires adequate computing power from the experimental platform in order to cope, in real time, with the computationally demanding optimization. In addition, both these strategies heavily rely on model identification of the system dynamics, which is usually subject to uncertainties both in structure and parameters \cite{cardinale2012contextualizing}. We also recall here the strategy in \cite{sootla2016shaping}, where a pulse-shaping controller is proposed to switch the system between its stable states, and the control strategies presented in \cite{chambon2019new}, where a piecewise linear switched control is designed to either stabilize a simplified toggle switch model onto its unstable equilibrium or to switch between the stable equilibria. Literature on learning-based control of the toggle switch is sparse when compared to the literature on model-based approaches. To the best of our knowledge, the first and only work in this direction was reported in \cite{sootla2013toggling}, where fitted Q-learning is used to toggle the switch between its stable equilibria, and the later extension of the approach to track periodic references  reported in \cite{sootla2013periodic}.

\section{The control problem} \label{sec:statment}

\begin{figure}[t]
\centering
\includegraphics[width=0.9\columnwidth]{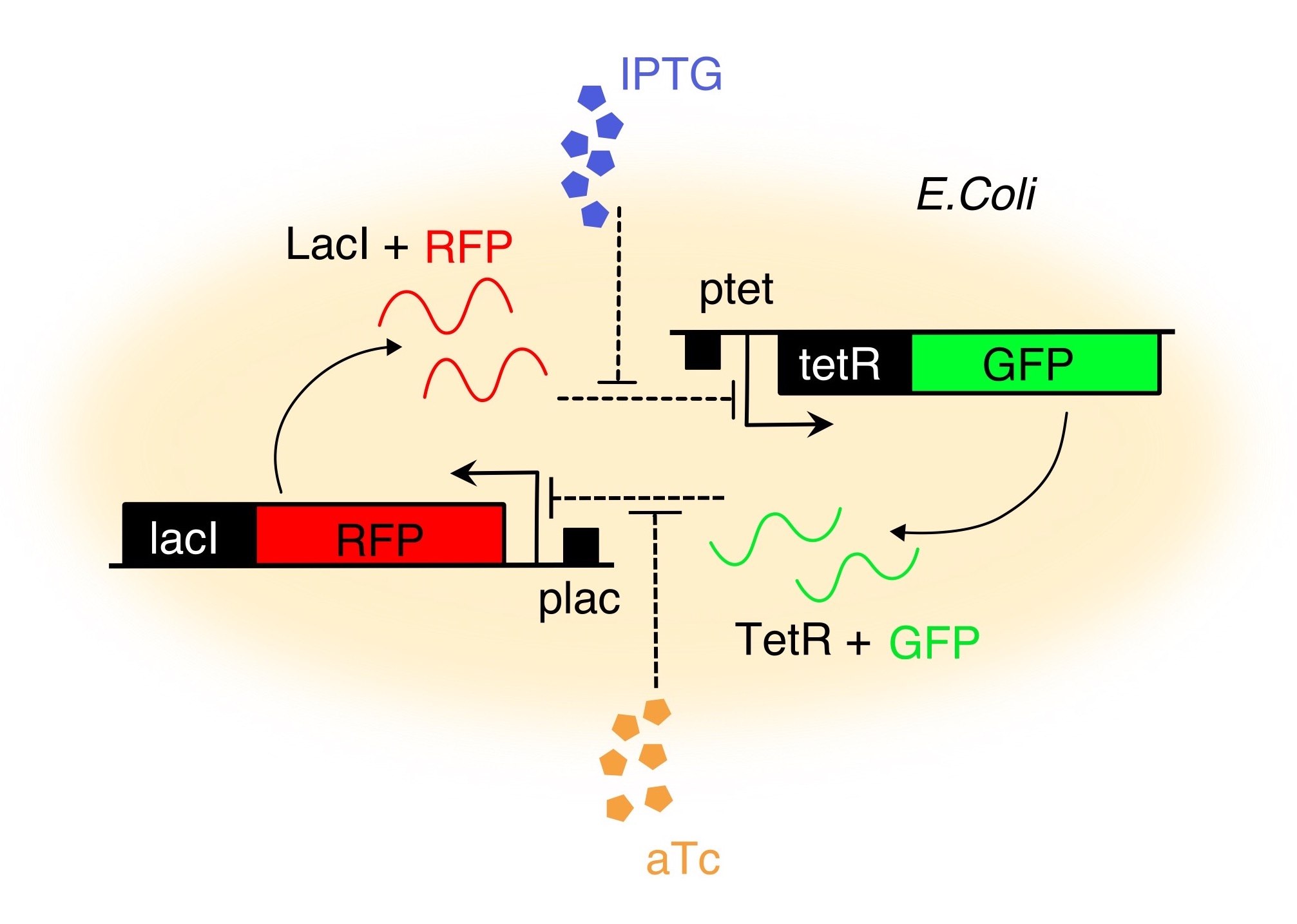}
\caption{Schematic of the genetic toggle switch circuit \cite{gardner2000construction}. Two genes, LacI and TetR, mutually repress each other; external inducers, IPTG and aTc, modulate the repression between the two proteins; the two genes are transcribed together with RFP and GFP, two fluorescent reporters that can be measured using microscopy.}
\label{fig:Toggle}
\end{figure}

The genetic toggle switch is a gene regulatory network first engineered in \textit{E.coli} in the early 2000 \cite{gardner2000construction}. It is constructed around two repressors, LacI and TetR, each inhibithing the other expression (Fig. \ref{fig:Toggle}). The double inhibition chain enables the creation of hysteresis, making this circuit a simple implementation of a binary memory element. 
From a dynamical system viewpoint, this genetic pathway is a bistable dynamical system, exhibiting at steady-state either of two stable phenotypes where one of the repressors is fully expressed and the other is scarcely present. 
The steady state phenotype exhibited can be modified by laboratory interventions that induce a transition between the two stable states.
Specifically, by adding two chemical inducers that diffuse through the cell membrane, i.e. anhydrotetracycline (aTc) and isopropyl-$\beta$-D-thiogalactoside (IPTG), it is possible to sequestrate TetR and LacI, respectively, relieving their inhibition on the competing repressor and toggling the system state.

\begin{figure}[!b]
\centering
\includegraphics[width=0.7\columnwidth]{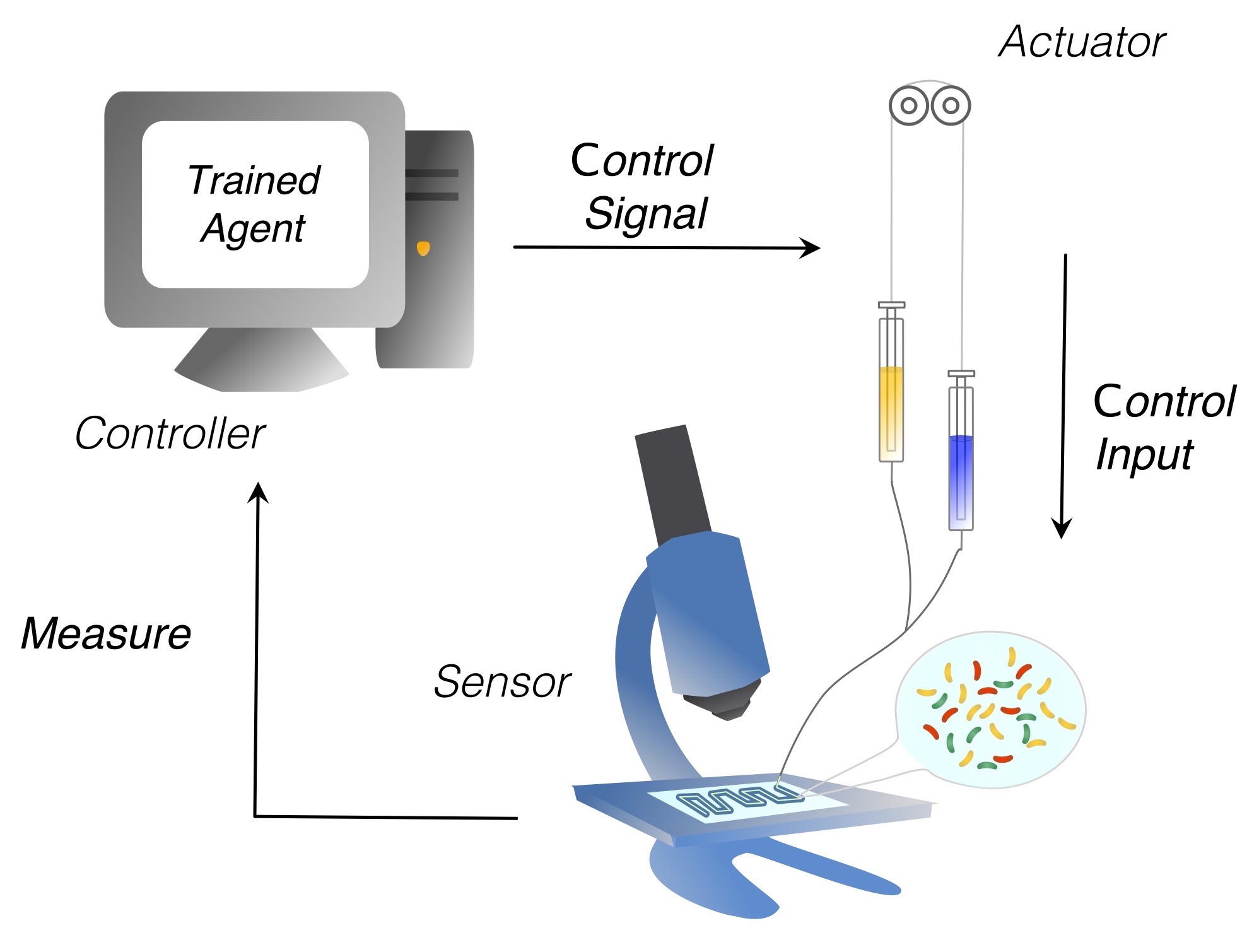}
\caption{Experimental set-up. A microfluidic device hosts a population of \textit{E.coli} endowed with a genetic toggle switch.  A microscope measures the average fluorescence levels of the reporter proteins RFP and GFP.  This information is fed to the trained artificial agent that computes the control input. The control signal is delivered via two motorized syringes.}
\label{fig:ControlLoop}
\end{figure}

In this work we consider the problem of balancing the genetic toggle switch around its unstable equilibrium by modulating the concentration of the inducers in the growth environment.
We address this problem by learning the policy via a model-free reinforcement learning algorithm. In our design, we explicitly account for biological and technological constraints of a real experimental set-up \cite{fiore2016vivo,ferry2011microfluidics}, schematically shown in Fig.~\ref{fig:ControlLoop}. As shown in the figure, the external control loop is implemented via a combination of microfluidics and inverted microscopy, through which it is possible to measure the fluorescence reporters expression level and to dynamically modify the composition of the growth medium. This experimental set-up, also described in \cite{fiore2016vivo,ferry2011microfluidics}, introduces  constraints on the time interval we can use to sample the state of the toggle switch and on the structure of the control input we can use to feed the cells. Namely,  constraints are:
\begin{description}
    \item[$\B{C1}$] The state of each cell can be sampled up to once every 5 min, to avoid excessive phototossicity;
    \item[$\B{C2}$] The concentration of the inducer molecules (i.e. the control inputs) can only be changed once every 15 min to limit osmotic stress on the cells;
    \item[$\B{C3}$] The sum of the inputs needs to be a convex combination of the maximum levels of inducers used in the  reservoirs. This is due to the specific implementation of the microfluidic device of choice.
\end{description}
%

\section{Mathematical Model} \label{sec:models}

In what follows, we denote sets with calligraphic capital characters, vectors with bold letters and random variables with capital letters. We denote by $\mathbb{R}$ and $\mathbb{R}_{\geq 0}$ the set of real and non-negative real numbers, respectively. Letting $\mathbf{v}\in\mathbb{R}^N$, then $\text{diag}(\mathbf{v})$ denotes the $\BB{R}^{N\times N}$ diagonal matrix having $\mathbf{v}$ on its main diagonal. All the quantities with the subscript $k$ refer to a discrete time domain. 

\subsection{Deterministic modeling}
We introduce  the mathematical model for the toggle switch that we will use later on for the \emph{in-silico} validations.
%
%
Letting  $x_1 \in \BB{R}_{\ge 0}$ be the concentration of the mRNA associated with the lacI gene ($\R{mRNA_{LacI}}$), $x_2 \in \BB{R}_{\ge 0}$ the concentration of the mRNA coding for TetR ($\R{mRNA_{TetR}}$), $x_3 \in \BB{R}_{\ge 0}$  the concentration of LacI, and $x_4 \in \BB{R}_{\ge 0}$ the concentration of TetR, 
using the pseudo reactions from \cite{lugagne2017balancing}, the dynamics of the system can be described via the following set of ODEs \cite{fiore2018analysis}:
\begin{subequations} \label{eq:full_det_model}
\begin{align}
\label{eq:LugagneOriginalMethods1}
& \frac{d}{dt} \, x_1 = k_\R{L}^\R{m0} + k^{\R{m}}_\R{L}h_{\R{LacI}}(x_4,v_1) -g_\R{L}^\R{m} \, {x_1}\\
\label{eq:LugagneOriginalMethods2}
& \frac{d}{dt} \, x_2 = k_\R{T}^\R{m0} + k^{\R{m}}_\R{T}h_{\R{TetR}}(x_3,v_2) - g_\R{T}^\R{m} \, {x_2}\\
\label{eq:LugagneOriginalMethods3}
& \frac{d}{dt} \, x_3= k_\R{L}^\R{p} \, {x_1} - g_\R{L}^\R{p} \, x_3\\
\label{eq:LugagneOriginalMethods4}
& \frac{d}{dt} \, x_4 = k_\R{T}^\R{p} \, {x_2} - g_\R{T}^\R{p} \, x_4
\end{align}
\end{subequations}
where $k_\R{L/T}^\R{m0}$, $k_\R{L/T}^\R{m}$, $k_\R{L/T}^\R{p}$, $g_\R{L/T}^\R{m}$, $g_\R{L/T}^\R{p}$ are basal transcription, maximal transcription, translation, mRNA degradation and protein degradation rates respectively, while  $v_1 \in \BB{R}_{\ge 0}$ and $v_2 \in \BB{R}_{\ge 0}$ are two inputs. Physically, these are the intra-cellular concentrations of aTc and IPTG, respectively. 
Note that $v_1$ and $v_2$ cannot be directly manipulated as we can only modify the concentration of the inducers in the culture media, setting their extra-cellular concentrations, say $u_1 \in \BB{R}_{\ge 0}$ and $u_2 \in \BB{R}_{\ge 0}$. 
Extra-cellular and intra-cellular inducer concentrations are linked together through the diffusion dynamics across the membrane. Specifically, as in \cite{lugagne2017balancing}, we complemented the model in \eqref{eq:full_det_model} with further equations describing diffusion:
\begin{subequations} \label{eq:diffusion}
\begin{align}
\label{eq:LugagneOriginalMethods5}
    \frac{d}{dt} \, v_1 & = \begin{cases}
    k^{in}_{aTc} (u_1 -v_1) \quad \text{if } u_1>v_1 \\
    k^{out}_{aTc} (u_1 -v_1) \quad \text{if } u_1\leq v_1
    \end{cases}\\
\label{eq:LugagneOriginalMethods6}
    \frac{d}{dt} \, v_2 & = \begin{cases} 
    k^{in}_{IPTG} (u_2 -v_2) \quad \text{if } u_2>v_2 \\
    k^{out}_{IPTG} (u_2 -v_2) \quad \text{if } u_2\leq v_2
    \end{cases}
\end{align}
\end{subequations}

The functions $h_{\R{LacI}}(x_4,v_1)$ and $h_{\R{TetR}}(x_3,v_2)$ in \eqref{eq:full_det_model} are Hill functions modeling the inhibition performed by LacI ($x_3$) and TetR ($x_4$). These are defined as
\begin{subequations} \label{eq:Hill_fcns_1}
\begin{align} 
    \label{eq:Hill_LacL}
    h_{\R{LacI}}(x_4, v_1) & = \frac{1}{1+\left(\frac{x_4}{\theta_{\R{TetR}}}\cdot h_{\R{aTc}}(v_1)\right)^{\eta_{\R{TetR}}}}\\
    \label{eq:Hill_TetR}
    h_{\R{TetR}}(x_3, v_2) & = \frac{1}{1+\left(\frac{x_3}{\theta_{\R{LacI}}}\cdot h_{\R{IPTG}}(v_2)\right)^{\eta_{\R{LacI}}}}
\end{align}
\end{subequations}
where $\theta_{\R{LacI}}$, $\theta_{\R{TetR}}$, $\eta_{\R{LacI}}$ and $\eta_{\R{TetR}}$ are constant positive parameters describing the dissociation constants and the hill coefficients, respectively.  In \eqref{eq:Hill_fcns_1} there are two other Hill functions modeling the repression relief actuated by aTc ($v_1$) and IPTG ($v_2$)
\begin{subequations} \label{eq:Hill_fcns_2}
\begin{align} 
    \label{eq:Hill_aTc}
    h_{\R{aTc}}(v_1) & = \frac{1}{1+\left(\frac{v_1}{\theta_\R{aTc}} \right)^{\eta_\R{aTc}}}\\
    \label{eq:Hill_IPTG}
    h_{\R{IPTG}}(v_2) & = \frac{1}{1+\left(\frac{v_2}{\theta_\R{IPTG}} \right)^{\eta_\R{IPTG}}}
\end{align}
\end{subequations}
where parameters $\theta_{\R{aTc}}$, $\theta_{\R{IPTG}}$, $\eta_{\R{aTc}}$, $\eta_{\R{IPTG}}$ have analogous meanings to those in \eqref{eq:Hill_fcns_1}.

\subsection{Stochastic model}
\label{sec:stochastic_model_TS}
In our \emph{in-silico} validations, we also consider the stochastic version of \eqref{eq:full_det_model} obtained from the Chemical Master Equation \cite{gillespie1992rigorous,salzano2019ratiometric,lugagne2017balancing}.  Specifically, following the arguments therein, we let $S$ be the stoichiometric matrix of the toggle switch given by
\begin{equation}
   S =\left[\begin{array}{cccccccc}
    1&0&0&0&-1&0&0&0 \\
    0&1&0&0&0&-1&0&0\\
    0&0&1&0&0&0&-1&0\\
    0&0&0&1&0&0&0&-1
    \end{array} \right]
\end{equation}
and $\mathbf{a}(\cdot)$ be the propensity vector which, for the toggle switch, is given by
\begin{equation} \label{eq:a_array}
\begin{split}
a=& [k^{\R{m0}}_L + k_\R{L}^\R{m} h_{\R{LacI}}(x_4,v_1),k^{\R{m0}}_\R{T} + k_\R{T}^\R{m} h_{\R{TetR}}(x_3,v_2),\\
&k_\R{L}^\R{p} x_1, k_\R{T}^\R{p} x_2,g_\R{L}^\R{m} x_1,g_\R{T}^\R{m} x_2,g^\R{p}_\R{L} x_3,g_\R{T}^\R{p} x_4]^T
\end{split}
\end{equation}
Then, the stochastic dynamics of the toggle switch can be given as
\begin{equation}
\label{sde}
d\mathbf{X}(t)= S\cdot \mathbf{a}(\mathbf{X}(t)) \cdot d t + S  \cdot \text{diag} (\sqrt{\mathbf{a}(\mathbf{X}(t))}) \cdot d\mathbf{W} 
\end{equation}
where $\mathbf{X(t)}= [X_1(t), X_2(t), X_3(t), X_4(t)]$ and  $d\B{W}\in \mathcal{R}^8$ are independent standard Wiener process increments \cite{lakatos2017stochastic}.
%
\begin{table} [!b]
	\centering
	\begin{tabular}{|c|c||c|c|}
		\hline 
		$ k_\R{L}^\R{m0} $ & $ 3.20e^{-2} $ mRNA $\min^{-1}$& $ g_\R{L}^\R{m}, g_\R{T}^\R{m} $ & $ 1.386e^{-1} $   \\
		\hline 
		$ k_\R{T}^\R{m0} $ & $ 1.19e^{-1} $ mRNA $\min^{-1}$& $ g_\R{L}^\R{p}, g_\R{T}^\R{p} $ & $ 1.65e^{-2} $  \\ 
		\hline 
		$ k_\R{L}^\R{m} $ & $ 8.30 $  mRNA $\min^{-1}$& $\theta_{\R{LacI}} $ & $ 31.94 $ \\ 
		\hline 
		$ k_\R{T}^\R{m} $ & $ 2.06 $ mRNA $\min^{-1}$& $\eta_{\R{LacI}} $ & $ 2.00 $\\
		\hline 
		$ k_\R{L}^\R{p} $ & $ 9.726e^{-1} $ a.u. $\text{mRNA}^{-1}$ $\min^{-1}$& $ x_2 $ & $ 30.00 $  \\ 
	    \hline 
		$ k_\R{T}^\R{p} $ & $ 9.726e^{-1}  $ a.u. $\text{mRNA}^{-1}$ $\min^{-1}$&  $ \eta_{\R{TetR}} $ & $ 2.00 $ \\ 
		\hline 
		$ k_{\R{aTc}}^{\R{in}} $ & $ 2.75e^{-2} $ $\min^{-1}$ & $ \theta_{\R{aTc}} $ & $ 11.65 $ \\
		\hline
		$ k_{\R{IPTG}}^{\R{in}} $ &  $ 1.62e^{-1}   $ $\min^{-1}$ & $ \eta_{\R{aTc}} $ & $ 2.00 $\\
		\hline
		$ k_{\R{aTc}}^{\R{out}} $ & $ 2.00e^{-2}  $ $\min^{-1}$ & $ \theta_{\R{IPTG}} $ & $ 9.06e^{-2}  $\\
		\hline
		$ k_{\R{IPTG}}^{\R{out}} $ &  $ 1.11e^{-1}   $ $\min^{-1}$ & $ \eta_{\R{IPTG}} $ & $ 2.00 $\\
		\hline
	\end{tabular}  
	\caption{Value of the parameters of the cell population models. Parameters taken from \cite{lugagne2017balancing}.}
	\label{tab:par}
\end{table}
%
\section{Training approach}
The intrinsic stochasticity of the biochemical reactions involved in the toggle switch makes model-based control approaches not always viable as they might lack robustness during \emph{in-vivo} validation \cite{lugagne2017balancing}. For this reason, we propose the use of a Q-learning algorithm (QL) in its classical tabular version to train an artificial agent using a standard $\epsilon$-greedy policy \cite{watkins1992q}. Differently from \cite{sootla2013toggling}, the goal here is to stabilize the switch onto its unstable equilibrium rather than toggling it from one stable equilibrium to the other.

%
\subsection{Control training approach}
To overcome the challenge of limited \emph{in-vivo} experiments that can be carried out to train the controller, we carry out the training using synthetic data generated via a simplified model of the switch that can be obtained from \eqref{eq:full_det_model} by using time scale separation, see \cite{fiore2018analysis} for its derivation.
%
%
%
%
%
\begin{figure}[!t]
\centering
\begin{tikzpicture}
  \node (img1)  {\includegraphics[width=\columnwidth]{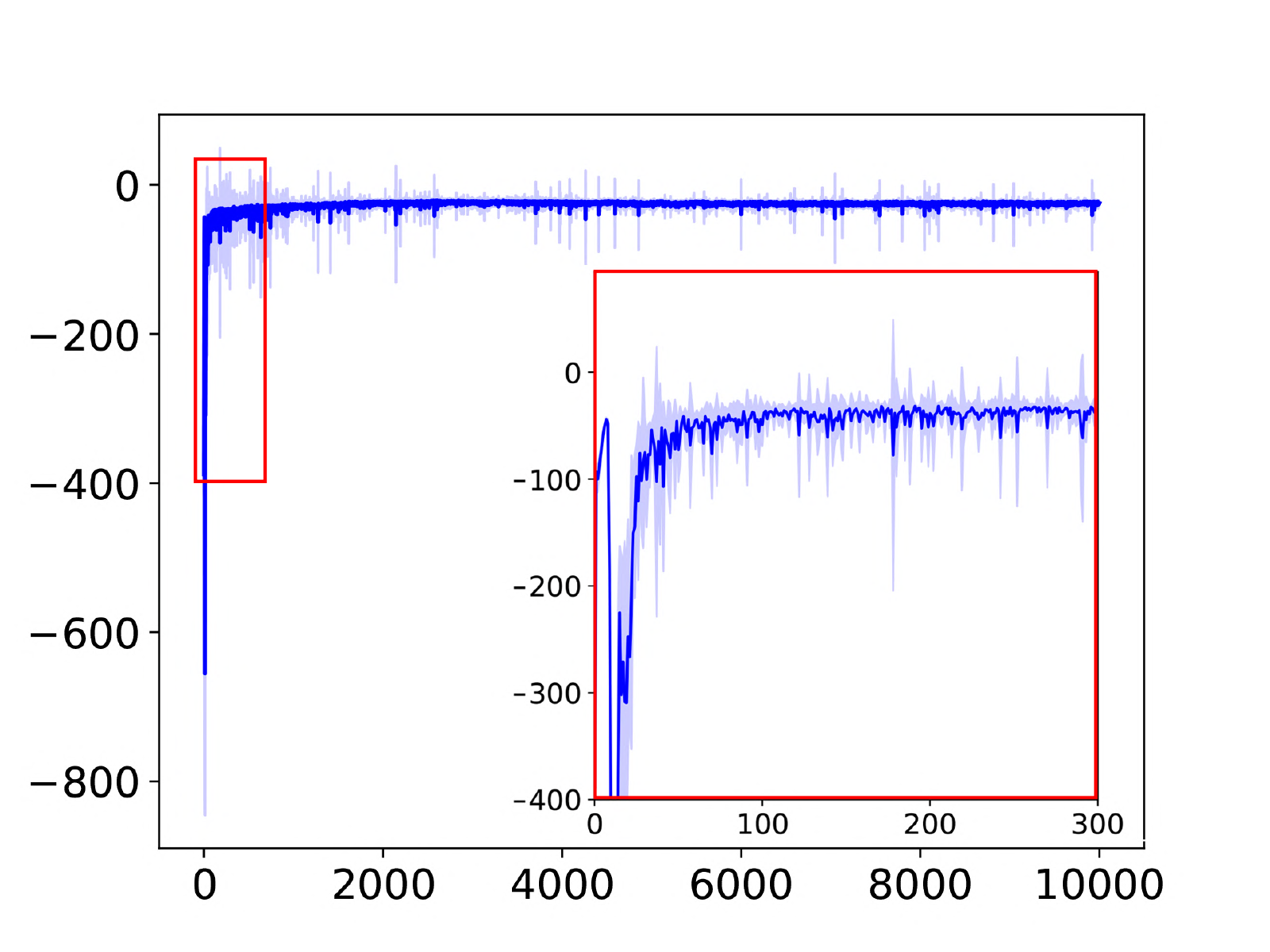}};
  \node[below=of img1, node distance=0cm, yshift=1cm,font=\color{black}] {episodes};
\end{tikzpicture}
\caption{Cumulative reward obtained over the $E=10000$ episodes. Training results are depicted in terms of average value (blue solid line) and standard deviation (blue shaded area) calculated on the $S=10$ realizations of the training process. In the red box it is shows the transient of the cumulative reward obtained during the first $300$ episodes. }
\label{fig:CumRew} 
\end{figure}
In particular we adopt the simplified adimensional model from \cite{fiore2018analysis}:
\begin{subequations} \label{eq:simple_model}
\begin{align}
\frac{dz_1}{dt'}& = k^0_1 + \frac{k_1}{1+z_2^2\cdot f_{\R{aTc}}(v_1(t'/g^p))} - z_1\\
\frac{dz_2}{dt'} & = k^0_2 + \frac{k_2}{1+z_1^2\cdot f_{\R{IPTG}}(v_2(t'/g^p))}  - z_2
\end{align}
\end{subequations}
where 
\begin{equation} \label{eq:var_change}
    z_1=\frac{x_3}{\theta_{\R{LacI}}},\quad z_2=\frac{x_4}{\theta_{\R{TetR}}}, \quad
    t'=g^\R{p}\, t,
\end{equation}
with $k^0_1=(k_\R{L}^\R{m0}\,k_\R{L}^\R{p})\big /(g_\R{L}^\R{m}\theta_{\R{LacI}} g^\R{p})$, $k^0_2  =(k_\R{T}^\R{m0}k_\R{T}^\R{p}\big/(g_\R{T}^\R{m}\theta_{\R{TetR}}g^\R{p})$, $k_2=(k_\R{L}^\R{m}k_\R{L}^\R{p})\big/(g_\R{L}^\R{m}\theta_{\R{LacI}}g^\R{p})$ and $k_2=(k_\R{T}^\R{m}k_\R{T}^\R{p})\big/(g_\R{T}^\R{m}\theta_{\R{TetR}}g^\R{p})$ being positive coefficients. The nonlinear function $f_{\R{aTc}}(\cdot)$ models the static relationship between the repressor protein LacI and its inducer molecule aTc, while $f_{\R{IPTG}}(\cdot)$ models the static relationship between the repressor protein TetR and its inducer molecule IPTG. Formally, these are defined as
\begin{equation*}
f_{\R{aTc}}(v_1) = \frac{1}{\big(1+v_1\big)^{\eta_{\R{TetR}}}},\quad
f_{\R{IPTG}}(v_2) = \ \frac{1}{\big(1+v_2\big)^{\eta_{\R{LacI}}}}.
\end{equation*}
The values of all the parameters used in the \emph{in-silico} experiments are reported in Table \ref{tab:par}.
\subsection{Training Results}
\label{fig:deterministic_simple}

Assuming the diffusion of the inducer molecules through the cell membrane to be faster than the  biochemical processes hosted in the cell, during training we neglected \eqref{eq:diffusion} by setting $u_1= v_1$ and $u_2=v_2$. Such an assumption will be later removed when applying the learnt policy to the more realistic toggle switch model of interest. Hence, providing further testing of whether the \textit{sim-to-real} gap can be effectively bridged.
%
{In our design, we fulfilled constraints $\B{C1}$ and $\B{C2}$ (see Sec. \ref{sec:statment}) by allowing the learning algorithm to measure the state of the system and modify the control inputs only once every $15$ minutes. This} corresponds to setting the sampling time of the virtual agent $T_s = g^p \cdot 15$ in the dimensionless model \eqref{eq:var_change}. {Finally, we fulfilled constraint $\B{C3}$ by enforcing the following conditions on $u_1$ and $u_2$:
\begin{equation}
    u_1 = \phi u_{1,\R{max}}, \quad u_2 = (1-\phi) u_{2,\R{max}}
\end{equation}
The QL was then used to learn the policy adjusting $\phi \in [0,1]$ (rather than for $u_1$ and $u_2$).} In the above expression $u_{1,\R{max}}= 35$, and $u_{2,\R{max}}= 0.35$ are the maximum values allowed for for the inducers' concentrations.

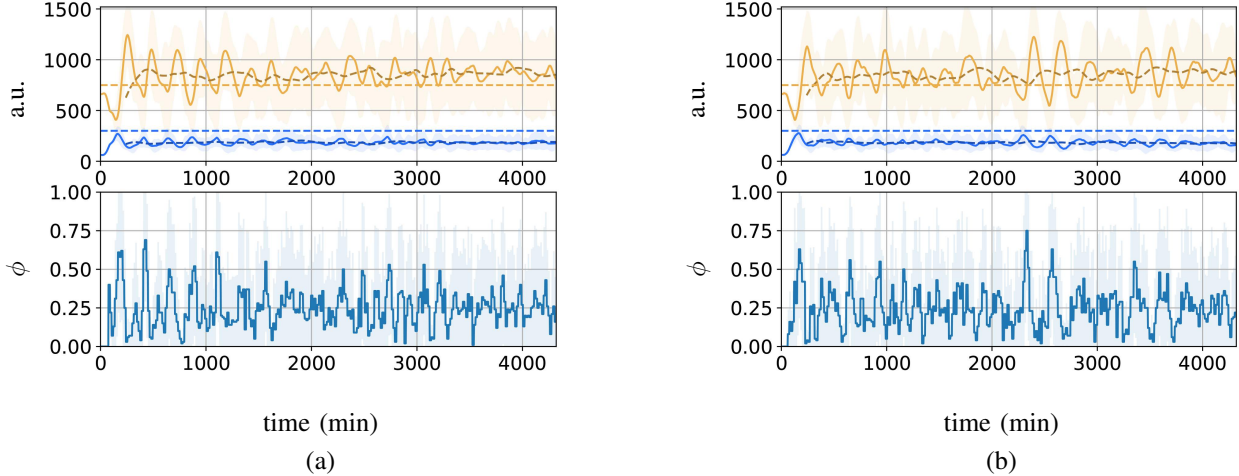
\begin{figure*}[t]

\begin{tikzpicture}
  \node (img1)  {\includegraphics[scale = 0.48]{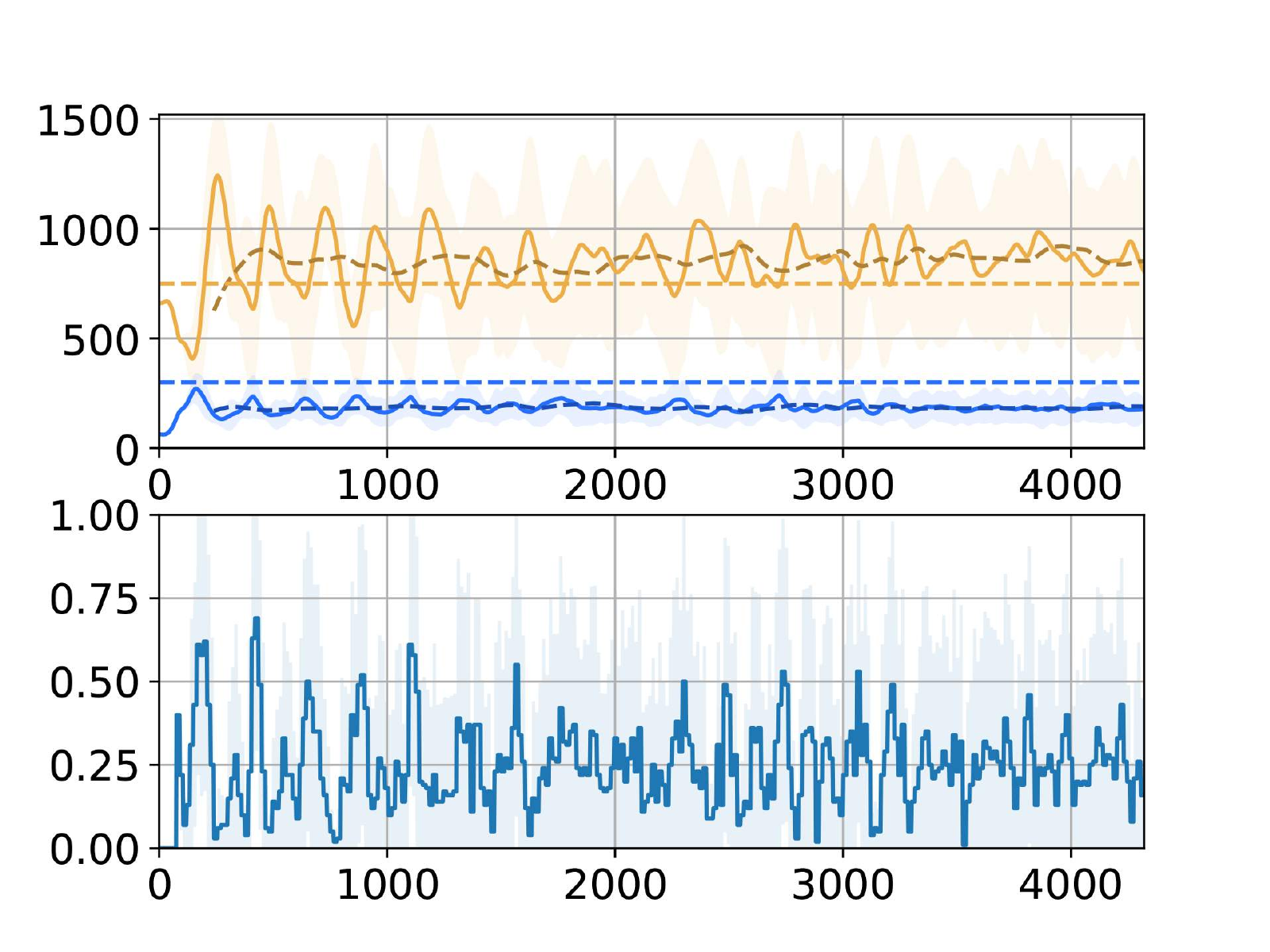}};
  \node[below=of img1, node distance=0.0cm, yshift=0.5cm,font=\color{black}] {(a)};
  \node[below=of img1, node distance=0cm, yshift=1cm,font=\color{black}] {time (min)};
  \node[left=of img1, node distance=0cm, rotate=90, anchor=center,xshift=-1.26cm ,yshift=-1.0cm,font=\color{black}] {$\phi$};
  \node[left=of img1, node distance=0cm, rotate=90, anchor=center,xshift=+1.0cm ,yshift=-1.0cm,font=\color{black}] {a.u.};
  \node[right=of img1] (img2)  {\includegraphics[scale = 0.48]{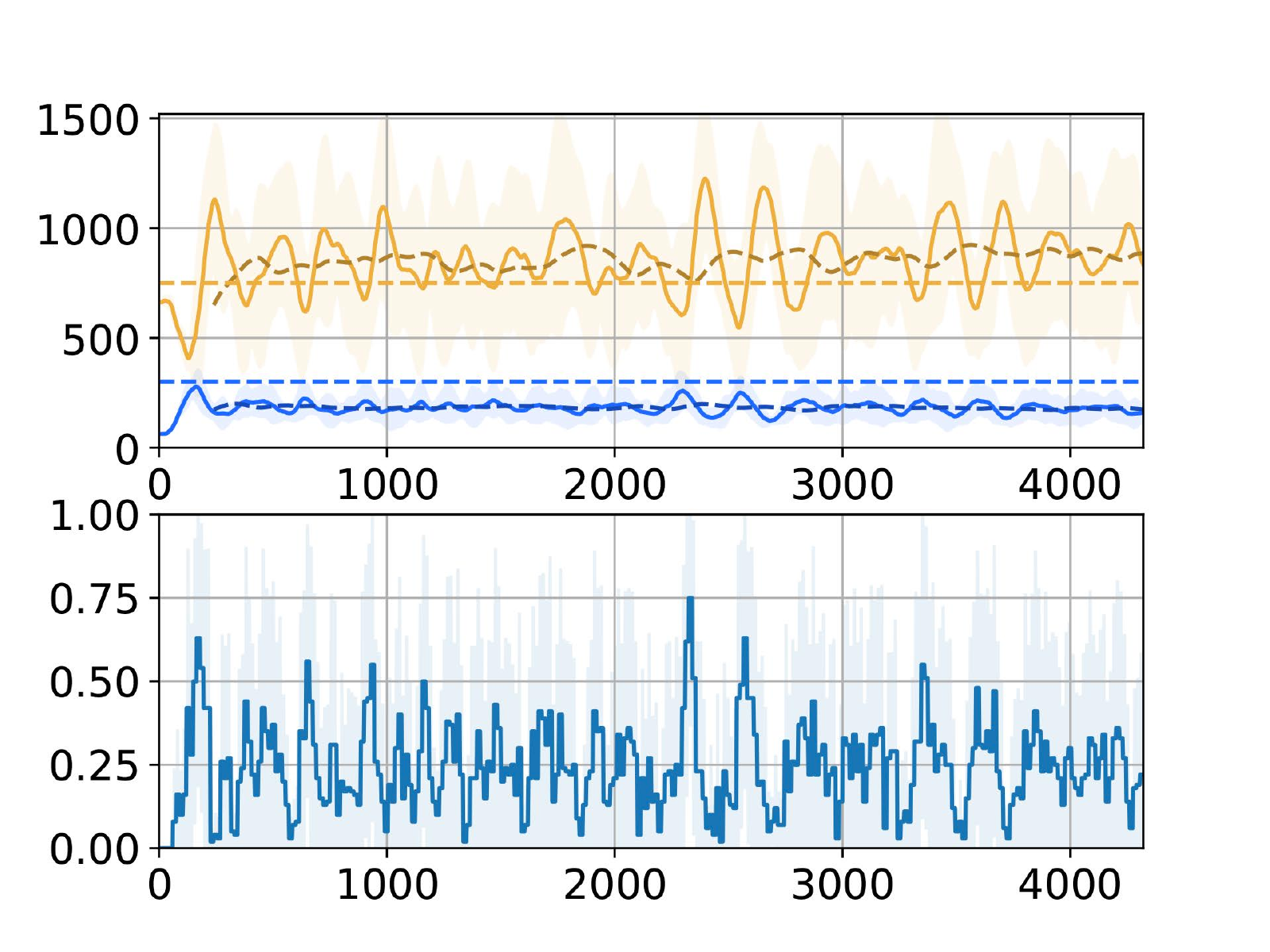}};
  \node[below=of img2, node distance=0.0cm, yshift=0.5cm,font=\color{black}] {(b)};
  \node[below=of img2, node distance=0cm, yshift=1cm,font=\color{black}]  {time (min)};
  \node[left=of img2, node distance=0cm, rotate=90, anchor=center,xshift=-1.26cm ,yshift=-1.0cm,font=\color{black}] {$\phi$};
  \node[left=of img2, node distance=0cm, rotate=90, anchor=center,xshift=+1.0cm ,yshift=-1.0cm,font=\color{black}] {a.u.};
\end{tikzpicture}
\caption{Simulation results carried on the toggle switch full dynamics in its deterministic (a) and stochastic (b) form. Top panels depict the time evolution of the system state while bottom panels depict the control input $\phi$ using the QL based controller. Solid lines represent the average evolution of states on inputs while the shaded areas represent the standard deviation obtained over $S=10$ realizations of the validation process. Light dashed lines indicate the set point of the experiment $(x_3, x_4)=[750,300]$ expressed in adimensional units (a.u.). Dark dashed lines indicate the average state trajectory evaluated with a moving window of a period of $240$ min.}
\label{fig:deterministic}
\end{figure*}

For the sake of comparison, we assumed as in \cite{guarino2020balancing} that the goal is to stabilize the system around the desired equilibrium point $\B{z}_{\R{ref}} = [23.48,10.00]^T$. In addition, we discretized the action-state space $\{\B{z_k},\phi_k\}$ as follows. The action space was sampled in 11 equally spaced values, while the states were discretized non-uniformly in the region $(z_1, z_2) \in [0,150] \times [0,150]$ which was heuristically found to be large enough. 
Specifically, we discretized each state in the region $(z_1,z_2) \in [z_{1,\R{ref}}-3, z_{1,\R{ref}}-3] \times  [z_{2,\R{ref}}-3, z_{2,\R{ref}}-3]$ with a step of $0.5$, while using a discretization step of $1.5$ elsewhere. By doing so, the artificial agent has a finer precision around the regulation point $\B{z_{\R{ref}}}$, reducing the steady state error.
%
As reward function, we selected the quadratic function
\begin{equation}
    r(\B{z}_k) = -\bigg( \bigg( \frac{z_{1,\R{ref}} - z_{1,k}}{z_{1,\R{ref}}}\bigg)^2 +\bigg( \frac{z_{2,\R{ref}} - z_{2,k}}{z_{2,\R{ref}}}\bigg)^2 \bigg) 
\end{equation}
%
We trained the virtual agent by running $S=10$ trials of $E=10000$ episodes each; an episode consisting of an \emph{in-silico} control experiment lasting $T=72$ hours. We set the learning rate as $\alpha = 0.8$, the probability of taking an exploratory move $\epsilon = 0.1$ and the discount factor $ \gamma = 0.9$. Results of the training are depicted in in Fig.~\ref{fig:CumRew} where the evolution of the cumulative reward is expressed in terms of mean and standard deviation over the $S=10$ realizations of the training process starting from the same initial conditions $\B{z_0} = [20.68,2.11]^T$.
We notice that the learning process reaches convergence on average within the first 200 episodes. For the sake of brevity, we do not show here the performance of the learnt control policy on the simplified model it was trained upon but choose to show later (see Fig. \ref{fig:deterministic}) its performance on the more realistic models presented in  Sec. \ref{sec:models}.

\section{In-silico validation on a realistic model}
We validated the performance of the control policy learnt on synthetic data generated by the simplified model by applying it to control the more realistic, higher-dimensional, deterministic and a  stochastic models presented in Sec. \ref{sec:models} which are used as proxies of a possible \emph{in-vivo} implementation of the toggle switch (as also done in \cite{lugagne2017balancing}).

\subsection{Metrics}
\label{sez:metrics}
We compared the performance of the control algorithm developed in this work with the PI-PWM and MPC strategies presented in \cite{guarino2020balancing}. Specifically, we quantified the effectiveness of each control strategy using the set of metrics introduced in \cite{fiore2016vivo,guarino2020balancing}. Namely, we computed the Integral Square Error (ISE), defined as
\begin{equation}
    \label{eq:ISE}
    ISE = \int_{t_0}^{T} \overline{e}(\tau)^2 d\tau,
\end{equation}
which measures the average transient and steady state performances. 
In addition, we evaluated the Integral Time-weighted Absolute Error (ITAE) defined as
\begin{equation}
\label{eq:ITAE}
    ITAE = \int_{t_0}^{T} \tau|\overline{e}(\tau)| d\tau
\end{equation}
which weighs the error more as the time progresses, making residual errors at steady state more relevant towards the computation of the control metric.
In Equations \eqref{eq:ISE}-\eqref{eq:ITAE}  $\bar{e}$ is computed as
\begin{equation}
\bar{e}(t) = \Bigl\lVert\big[\frac{\overline{x_3} -x_{3,\R{ref}}}{x_{3,\R{ref}}}; \frac{\overline{x_4} -x_{4,\R{ref}}}{x_{4,\R{ref}}}\big]\Bigl\rVert_2 
\end{equation}
with $\bar{x}_3(t)$ and $\bar{x}_4(t)$ being the moving averages of $x_3(t)$ and $x_4(t)$ over a window of width $t_w$, formally defined as 
\begin{equation*}
    \overline{x}_3(t) = \frac{1}{t_w}\int_{t-t_w}^{t} x_3(\tau) d\tau, \quad
    \overline{x}_4(t) = \frac{1}{t_w}\int_{t-t_w}^{t} x_4(\tau) d\tau.
\end{equation*}
Both metrics are evaluated from $t_0 = t_w$ up to the control horizon $T$.

\subsection{Deterministic experiments}
 Fig.~\ref{fig:deterministic}.a shows the control performance of the learnt control strategy when applied to the  realistic deterministic model \eqref{eq:full_det_model} showing the effectiveness of the proposed training strategy. Table \ref{table:2} shows a quantitative comparison between the strategy presented here with the MPC and PI-PWM previously described in \cite{guarino2020balancing}, when both are tested on \eqref{eq:full_det_model}. We see that the performance achieved by the QL based controller are comparable with the model-based ones. In particular, the model free QL is capable of outperforming the PI-PWM in terms of ISE, reducing the variability of the proteins during the transient.

\subsection{Stochastic experiments}
Next we test the robustness of the proposed approach by running a set of \emph{in-silico} experiments using the stochastic model of the toggle switch presented in Sec. \ref{sec:stochastic_model_TS} using a Euler-Maurayama scheme with $dt = 0.1$ for its efficient numerical implementation \cite{higham2001algorithmic}.  The outcome of the \emph{in-silico} experiments are reported in Fig.~\ref{fig:deterministic}.b. Table \ref{table:2} shows the quantitative comparison between QL, MPC and PI-PWM when applied to the stochastic model. Again, we see that the QL based controller shows comparable performance when contrasted with model-based ones for the control of the full model confirming the viability of a \textit{sim-to-real} paradigm in a biological setting. 

\section{Conclusions}
We investigated the problem of stabilizing the unstable equilibrium of a genetic toggle switch using via an external control approach based on machine learning. To overcome the data efficiency problem that would render the algorithm unfeasible for practical use in synthetic biology, we adopted and tested \emph{in-silico} the use of a \emph{sim-to-real} paradigm. That is, the policy was first learnt via training on a simplified model of the toggle switch and  it was then subsequently exploited to control a more realistic model of the switch parameterized from \emph{in-vivo} experiments. 

Our \emph{in-silico} experiments confirmed the viability of this approach suggesting its potential use for \emph{in-vivo} control implementations. This represents a crucial step towards the deployment of learning algorithms to control synthetic biological circuits. Ongoing research is aimed at testing the proposed strategy \emph{in-vivo} using the microfluidics platform described in \cite{menolascina2014vivo}.

\begin{table}[t]
\begin{center}
\begin{tabular}{|c c c c c|} 
 \hline
   & QL & QL & MPC & PI-PWM \\
   & \emph{quasi-steady} & complete  & & \\
   & \emph{state} assumption & model & & \\ 
 \hline
\multicolumn{5}{ |l| }{\textbf{Deterministic \emph{in-silico} experiments}} \\
 ISE  & $113.23$ & $767.04$ & $47.58$ & $876.71$\\
 ITAE & $1.51\R{E}^{+6} $ & $3.96\R{E}^{+6}$ & $0.81\R{E}^{+6} $ & $2.07\R{E}^{+6} $ \\
 \hline
\multicolumn{5}{ |l| }{\textbf{Stochastic \emph{in-silico} experiments}} \\
 ISE  &$116.86$ & $794.64$ & $178.50$ & $830.52$\\
 ITAE & $1.53\R{E}^{+6} $ & $4.07\R{E}^{+6}  $ & $1.98\R{E}^{+6} $ & $2.07\R{E}^{+6} $\\
 \hline
\end{tabular}
\caption{Control performance comparison via the metrics introduced in Sec.\ref{sez:metrics} for deterministic and stochastic experiments}
\label{table:2}
\end{center}
\end{table}

\printbibliography
\end{document}